\documentstyle[array,epsfig,twocolumn,pra,aps]{revtex}

\newcommand{\ep}{\epsilon}

\begin{document}
\draft
\wideabs{
  
  \title{Efficient quantum computation using coherent states}
  
  \author{H. Jeong
     and M. S. Kim
    }
  
  \address{School of Mathematics and Physics, Queen's University,
    Belfast BT7 1NN, United Kingdom 
    }
  
  \date{\today}  
  \maketitle

\begin{abstract}
  Universal quantum computation using optical coherent states is studied.
  A teleportation scheme for a coherent-state qubit is developed and
  applied to gate operations.  This scheme is shown to be robust to detection
  inefficiency.
\end{abstract}
}

\section{Introduction}

The theory of quantum computation promises to revolutionize the future
of computer technology  with merits in factoring large integers
\cite{Shor} and combinational searches \cite{Grover}.
In recent years, the physical implementation of a quantum computer
has been intensively studied.  
Quantum computing in optical systems has been
studied as one of several plausible models.  
Recently, Knill {\it et al.} suggested a scheme for efficient quantum
computation with linear optics \cite{Knill}.

A coherent field is a fundamental tool in quantum optics and linear
superposition of two coherent states is considered one of the
realizable mesoscopic quantum systems \cite{Yurke}. In particular,
Cochrane {\em et al.} \cite{Cochrane} showed how logical qubits can be
implemented using even and odd coherent superposition states which are
defined as $N_\pm^0(|\alpha\rangle\pm|-\alpha\rangle)$ with
$|\alpha\rangle$ and $|-\alpha\rangle$ representing coherent states of
$\pi$ phase difference and $N_\pm^0$ being the normalization factors.  The
two superposition states form orthogonal bases in two-dimensional
Hilbert space and they can be discriminated by photon measurement
\cite{Buzek}.  There were some proposals to entangle such the logical
qubits with atomic states \cite{Munro}.  One drawback of using even
and odd cat states as a logical qubit basis for quantum computation is
that they are extremely sensitive to photon loss and detection
inefficiency.

In this paper, we present a method to implement
universal quantum computation using coherent states.
This proposal makes it possible to realize universal 
 quantum
computation
based on quantum teleportation \cite{Bennett93} which was shown to be
a useful tool in controlled gate operation \cite{GC}.  It is also found that
this scheme is robust to detection inefficiency.

\section{readout scheme and universal gate operations}

Let us consider two coherent states $|\alpha\rangle$ and
$|-\alpha\rangle$, where the coherent amplitude $\alpha$ is taken to
be real. The two coherent states are not orthogonal to each other but
their overlap $\langle\alpha|-\alpha\rangle=\mbox{e}^{-2\alpha^2}$
decreases exponentially with $\alpha$.  For example, when $\alpha$ is
as small as 3, the overlap is $\approx 10^{-8}$.  Throughout the
paper, the average photon number of the coherent state is assumed
around 10.  We identify the two coherent states of $\alpha$ as basis
states for a logical qubit:
\begin{equation}
\label{base}
|\alpha\rangle\rightarrow|0_L\rangle,~~~|-\alpha\rangle\rightarrow|1_L\rangle.
\end{equation}
A qubit state is then represented by $|\phi\rangle={\cal
  A}|\alpha\rangle+{\cal B}|-\alpha\rangle$ where the normalization
  condition is
\begin{eqnarray}
1&=&\langle\phi|\phi\rangle=|{\cal A}|^2+|{\cal
  B}|^2+({\cal AB^*+A^*B})\langle\alpha|-\alpha\rangle\nonumber\\
&\approx&|{\cal A}|^2+|{\cal B}|^2.
\label{nap}
\end{eqnarray}

\begin{figure}
\centerline{\scalebox{0.58}{\includegraphics{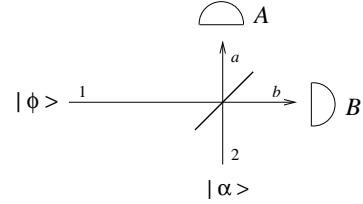}}}
\vspace{0.3cm}
\caption{Measurement scheme for $|\phi\rangle_1={\cal
    A}|\alpha\rangle_1+{\cal B}|-\alpha\rangle_1$ with a 50-50 beam
  splitter and auxiliary state $|\alpha\rangle_2$.
If detector $A$ registers any photon(s) while
  detector $B$ does not, the measurement outcome is $|\alpha\rangle$,
  {i.e.} $|0_L\rangle$.
  On the contrary, $A$ does not click while $B$ does, the measurement
  outcome is $|-\alpha\rangle$, {i.e.} $|1_L\rangle$. }
\label{fig:readout}
\end{figure}

Let us consider the readout of a qubit. The logical basis states, $|\alpha\rangle$ and
$|-\alpha\rangle$, can be discriminated by a simple measurement scheme
with a 50-50 beam splitter, an auxiliary coherent field of amplitude $\alpha$ and
two photodetectors as shown in Fig.~\ref{fig:readout}. At
the beam splitter, the input state $|\phi\rangle_1$ is superposed with
the auxiliary
state $|\alpha\rangle_2$ and gives the output
\begin{equation}
\label{eq:ax}
|\phi_R\rangle_{ab}={\cal A}|\sqrt{2}\alpha\rangle_a|0\rangle_b+{\cal B}|0\rangle_a|-\sqrt{2}\alpha\rangle_b.
\end{equation}
If detector $A$ registers any photon(s) while detector $B$ does not,
we know that $|\alpha\rangle$ is measured. On the contrary, if $A$
does not click while $B$ does, the measurement outcome is
$|-\alpha\rangle$.  Even though there is a non-zero probability of
failure $P_f=|_a\langle0|_b\langle0|\phi_R\rangle_{ab}|^2=|{\cal A}+{\cal
  B}|^2\mbox{e}^{-2\alpha^2}$ in which both of the detectors do not
register a photon, the failure is known from the result whenever it
occurs, and $P_f$ approaches to zero exponentially as $\alpha$
increases.

\begin{figure}
\centerline{\scalebox{0.37}{\includegraphics{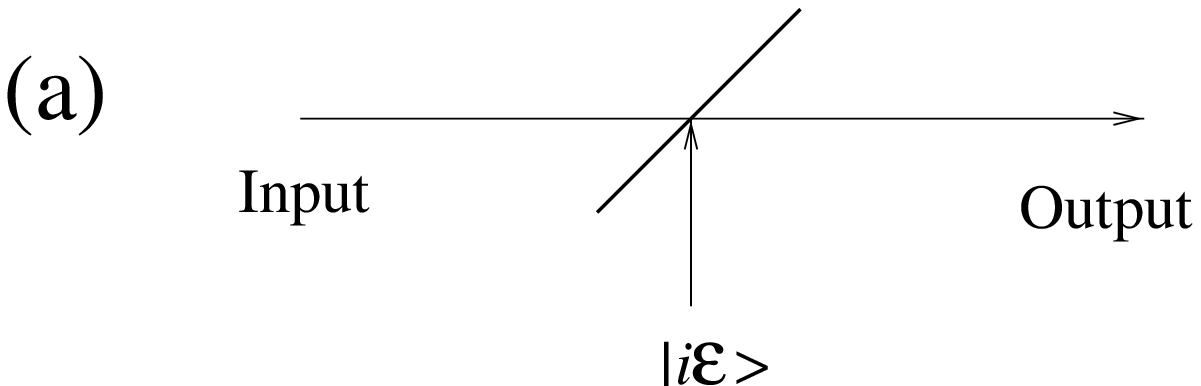}}}
\vspace{0.3cm}
\centerline{\scalebox{0.37}{\includegraphics{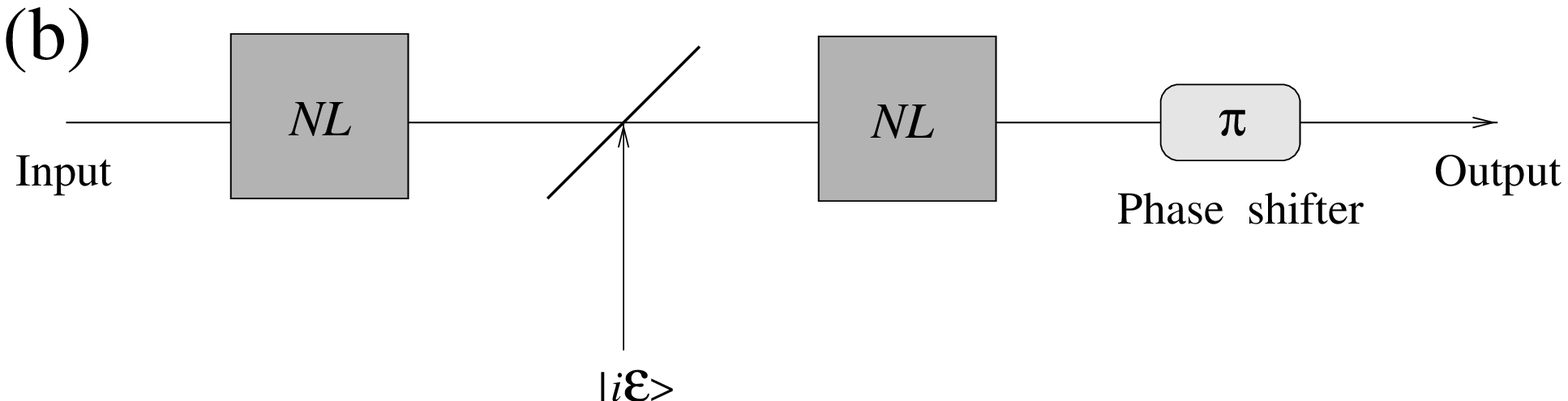}}}
\vspace{0.3cm}
\centerline{\scalebox{0.37}{\includegraphics{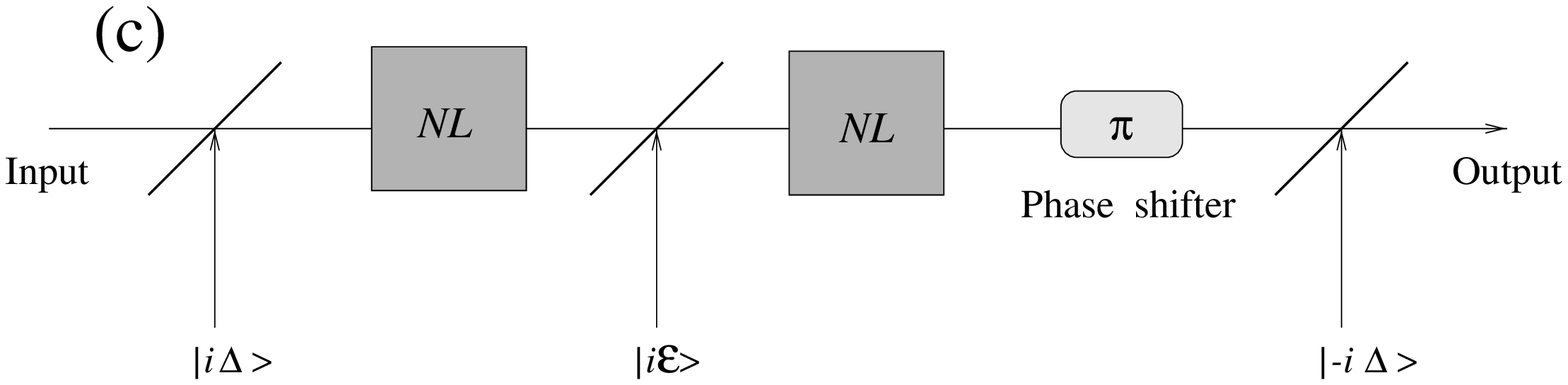}}}
\vspace{0.3cm}
\caption{1-bit rotation around the $z$ (a), $y$ (b), and $x$  axes (c)
  for a qubit state of coherent fields. $NL$ represents
  a nonlinear medium.  The
  transmission coefficient $T$ of the beam splitters is assumed to be
  close to unity. ${\cal E}$ corresponds to
  $\frac{\theta}{4\alpha\sqrt{1-T}}$, where $\theta$ is the required
  degree for a rotation and $\alpha$ is the coherent amplitude of the
  qubit state $|\phi\rangle$.  $\Delta=\frac{\pi}{8\alpha\sqrt{1-T}}$.
  Starting from a coherent state, an arbitrary qubit can be prepared
  up to a global phase using the above operations.  }
\label{fig:hp}
\end{figure}

\begin{figure}
\centerline{\scalebox{0.37}{\includegraphics{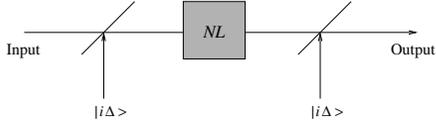}}}
\vspace{0.3cm}
\caption{Hadamard gate for a qubit state  $|\phi\rangle$ = ${\cal
    A}|\alpha\rangle$ + ${\cal B}|-\alpha\rangle$. The coherent field amplitude
  $i\Delta$ is $i\frac{\pi}{8\alpha\sqrt{1-T}}$ and the transmission
  coefficient $T$ of the beam splitters is close to unity.  The irrelevant
  global phase is neglected.  }
\label{fig:h}
\end{figure}

An arbitrary 1-bit rotation and a controlled-NOT (CNOT) gate for
two-qubit states form a set which satisfies all the requirements for a
universal gate operation.  For any SU(2) unitary operation, there is a
unique rotation $R(\theta,\phi,\eta)$ around the $x$, $y$ and $z$
axes.
 Cochrane {\em et al.} showed that the rotation around $x$ axis for even and odd coherent superposition states
can be realized using a interaction Hamiltonian $H_D=\hbar(\beta a^\dagger +\beta^*
a)$, where $\beta$ is the complex amplitude of the classical driving
force \cite{Cochrane}.
The evolution by this Hamiltonian corresponds to the displacement operator,
$D(\delta)=\exp(\delta a^\dag-\delta^* a)$, where $a$ and $a^\dag$ are
respectively annihilation and creation operators. 
In a similar way, $z$-rotation 
\begin{equation}
U_z(\theta/2)=\left(
\begin{array}{cc}
e^{i\theta/2} & 0   \\
0 & e^{-i\theta/2} 
\end{array}\right)
\end{equation}
for a logical qubit $|\phi\rangle$ can be obtained.  A coherent state
is a displaced vacuum $|\alpha\rangle=D(\alpha)|0\rangle$.  We know
that two displacement operators $D(\alpha)$ and $D(\delta)$ do not
commute but the product $D(\alpha)D(\delta)$ is simply
$D(\alpha+\delta)$ multiplied by a phase factor,
$\exp[(\alpha\delta^*-\alpha^*\delta)/2]$.  This phase factor plays a
role to rotate the logical qubit.  The action of displacement operator
$D(i\ep)$, where $\ep$ ($\ll \alpha$) is real, on the qubit $|\phi\rangle$
is the same as the $z$-rotation of the qubit by $U_z(2\alpha\ep)$.  We can
easily check their similarity by calculating the fidelity:
\begin{eqnarray}
\label{fidelity-rotation}
&&|\langle \phi|U_z^\dag(2\alpha\ep)D(i\ep)|\phi\rangle|^2\nonumber \\
&&~=e^{-\epsilon^2}\big\{|{\cal A}|^2+|{\cal
  B}|^2+e^{-2\alpha^2}({\cal AB^*}e^{-2i\alpha\epsilon}+{\cal A^*B}e^{2i\alpha\epsilon})
\big\}^2\nonumber\\
&&~\approx \exp[-\ep^2]\approx1.
\end{eqnarray}
Thus the rotation angle $\theta$ depends on $\alpha$ and $\ep$:
$\theta=4\alpha\ep$.
A small amount of $\ep$ suffices to make one cycle of rotation as
$\alpha$ is relatively large.
The displacement operation $D(i\ep)$ can be effectively performed using a beam splitter with the
transmission coefficient $T$ close to unity and a high-intensity coherent field of amplitude
$i{\cal E}$, where ${\cal E}$ is real, as shown in
Fig.~\ref{fig:hp}(a).  It is known that the effect of the beam
splitter is described by $D(i{\cal E}\sqrt{1-T})$ in the limit of
$T\rightarrow1$ and ${\cal E}\gg1$. (More rigorously the output state
becomes mixed but in the limit it can well be approximated to a pure
state as shown by one of the authors \cite{Kim}.)

To achieve any arbitrary 1-bit rotation, we need to operate
$U_x(\pi/4)$ and $U_x(-\pi/4)$ which are rotations by $\pi/2$ and
$-\pi/2$, respectively, around the $x$ axis.  We find that $U_x(\pi/4)$ can be
realized using a nonlinear medium.  Even though the efficiency of nonlinear interaction
can be a problem, there was an experimental
report for a successful measurement of giant Kerr nonlinearity \cite{Hau}.
 The anharmonic-oscillator
Hamiltonian of an amplitude-dispersive medium is
\cite{Yurke}
\begin{equation}
\label{eq:nonlinear}
{\cal H}_{NL}=\hbar\omega a^\dag a+\hbar\Omega(a^\dag a)^2,
\end{equation}
where $\omega$ is the frequency of the coherent field and $\Omega$ is the
strength of the anharmonic term.  When the interaction time $t$ in the
medium is $\pi/\Omega$, coherent states $|\alpha\rangle$ and
$|-\alpha\rangle$ evolve as follows:
\begin{eqnarray}
&&|\alpha\rangle
\longrightarrow\frac{e^{-i\pi/4}}{\sqrt{2}}(|\alpha\rangle+i|-\alpha\rangle),\\
&&|-\alpha\rangle
\longrightarrow\frac{e^{-i\pi/4}}{\sqrt{2}}(i|\alpha\rangle+|-\alpha\rangle).
\end{eqnarray}
This transformation corresponds to $U_x(\pi/4)$ up to a global phase
shift. The other rotation $U_x(-\pi/4)$ can be realized by applying a
phase shifter $P(\pi)$, which acts
$|\alpha\rangle\leftrightarrow|-\alpha\rangle$, after or before
$U_x(\pi/4)$ operation.  Note that $P(\pi)$ corresponds to
$\pi$-rotation around the $x$ axis, {\it i.e.} a 1-bit NOT gate.  The
other two required unitary operations $U_y(\phi/2)$ and $U_z(\eta/2)$
which correspond to rotations around the $y$ and $x$ axes can be
realized using the following identities \cite{Arfken}
\begin{eqnarray}
U_y(\phi/2)&=&U_x(-\pi/4)U_z(\phi/2)U_x(\pi/4), \\
U_x(\eta/2)&=&U_z(-\pi/4)U_y(\eta/2)U_z(\pi/4).
\end{eqnarray}
Therefore, any 1-bit rotation 
can be performed up to a 
global phase with beam splitters, nonlinear media, phase shifters and
auxiliary coherent light fields as shown in Fig.~\ref{fig:hp}.  As an
example, we can construct the Hadamard gate $H$ as
\begin{equation}
H=-U_z(\pi/4)U_x(\pi/4)U_z(\pi/4),
\end{equation}
which is shown in Fig~\ref{fig:h}.  Using these operations, any 1-qubit
state $|\phi\rangle={\cal A}|\alpha\rangle+{\cal B}|-\alpha\rangle$ with arbitrary ${\cal A}$
and ${\cal B}$ can be prepared up to a global phase from a coherent
state.

\begin{figure}
\centerline{\scalebox{0.5}{\includegraphics{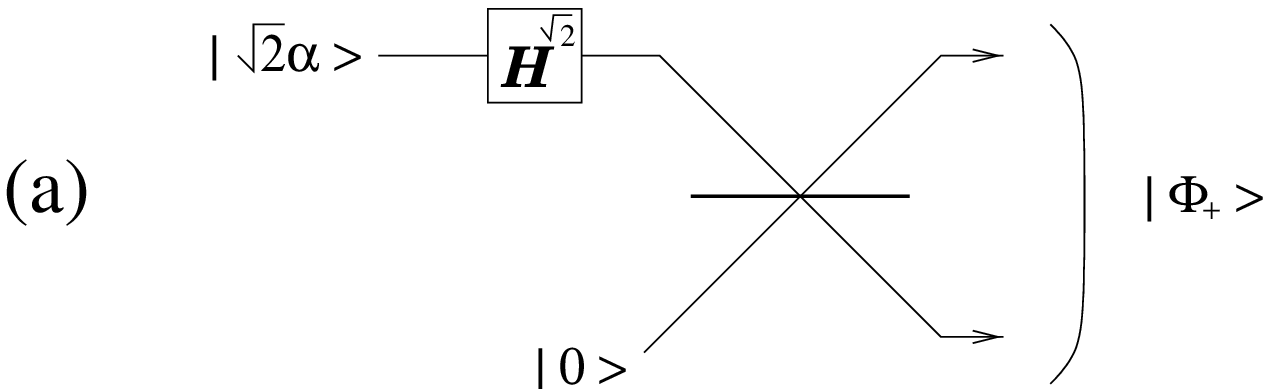}}}
\vspace{0.5cm}
\centerline{\scalebox{0.5}{\includegraphics{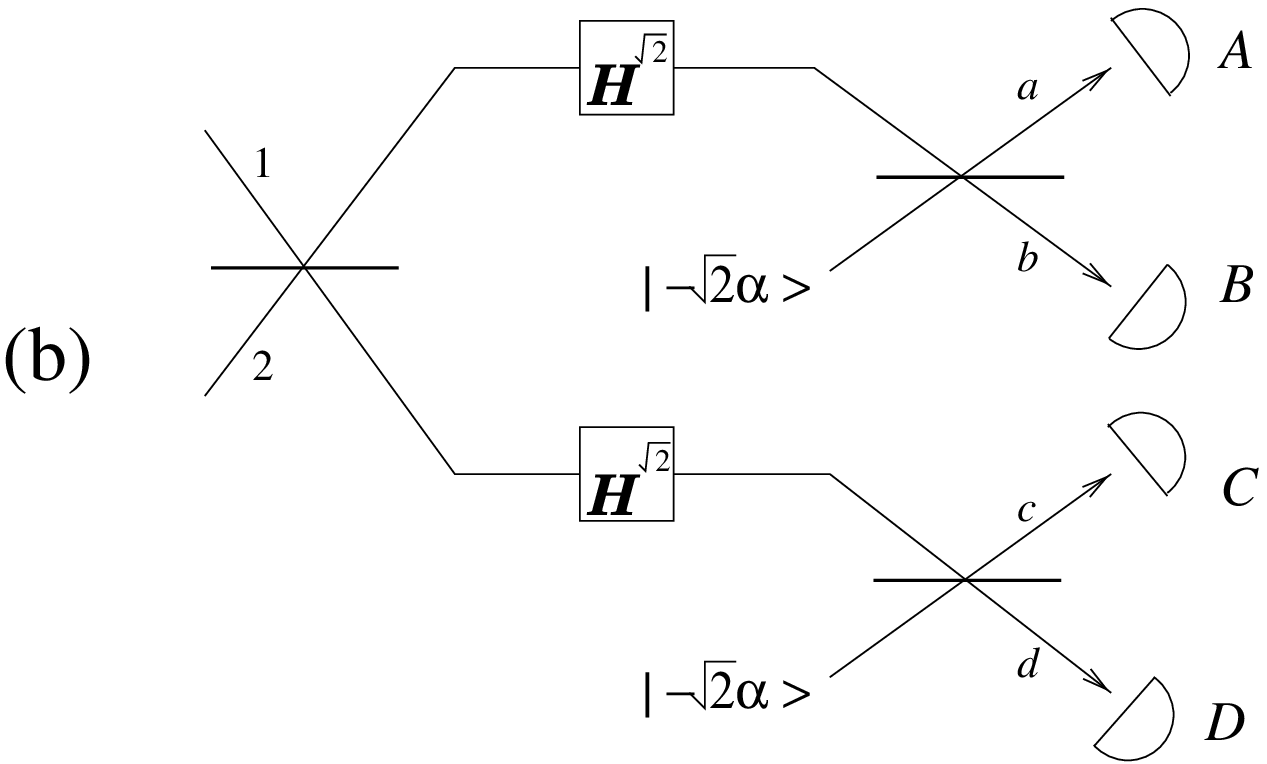}}}
\vspace{0.5cm}
\centerline{\scalebox{0.5}{\includegraphics{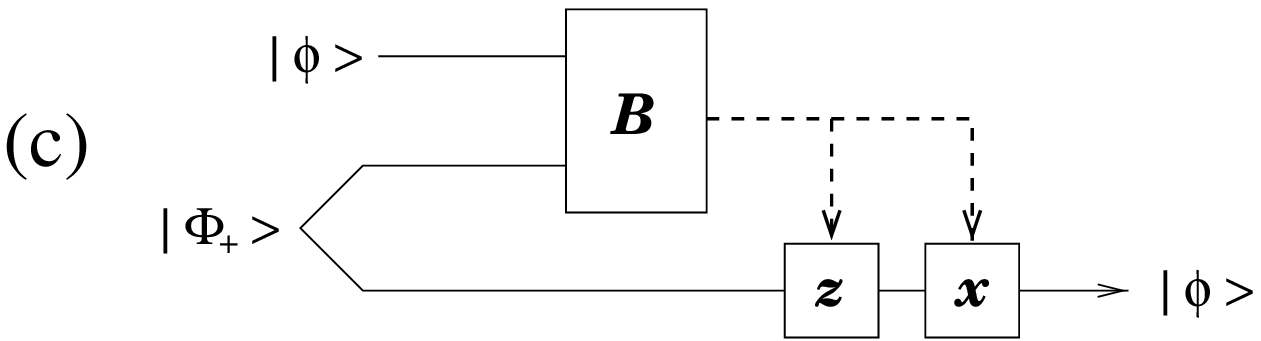}}}
\vspace{0.3cm}
\caption{Teleportation process for an unknown state $|\phi\rangle={\cal
    A}|\alpha\rangle+{\cal B}|-\alpha\rangle$.  $H^{\sqrt 2}$
  represent the Hadamard gate with an incident qubit state of coherent
  amplitudes $\pm\sqrt{2}\alpha$. $B$ represents the Bell measurement.
  $x$ and $z$ represent $\pi$ rotation around the $x$ and $z$ axes.
  (a) Generation of the quantum channel $|\Phi_+\rangle$. (b)
  Bell-state measurement with arbitrarily high precision. If detector
  $A$ does not click, the measurement outcome is $|\Phi_+\rangle$, and
  so on.  Only one of the four detectors does not detect any photon at
  a measurement event for $\alpha\gg1$.  (c) Scheme to teleport
  $|\phi\rangle$ via the entangled quantum channel $|\Phi_+\rangle$.
  The Pauli operations represented by $x$ and $z$ are performed
  according to the result of Bell measurement $B$.}
\label{fig:tc}
\end{figure}

For a universal gate operation, a CNOT gate is required besides 1-bit
rotation.  
It was found that the CNOT operation can be realized using a
teleportation protocol \cite{GC}.  For a superposition of coherent
states, quantum teleportation protocols have been suggested by
utilizing an entangled coherent state \cite{Enk00,JKL01} including an
entanglement purification scheme \cite{JKL01}.  However, the success
probability of this teleportation scheme is limited to less than 1/2
in practice and the required photon parity measurement is very
sensitive to detection 
inefficiency and photon loss as the parity
alternates by missing one photon. We suggest a teleportation protocol
as follows to circumvent those problems.

\begin{figure}
\centerline{\scalebox{0.5}{\includegraphics{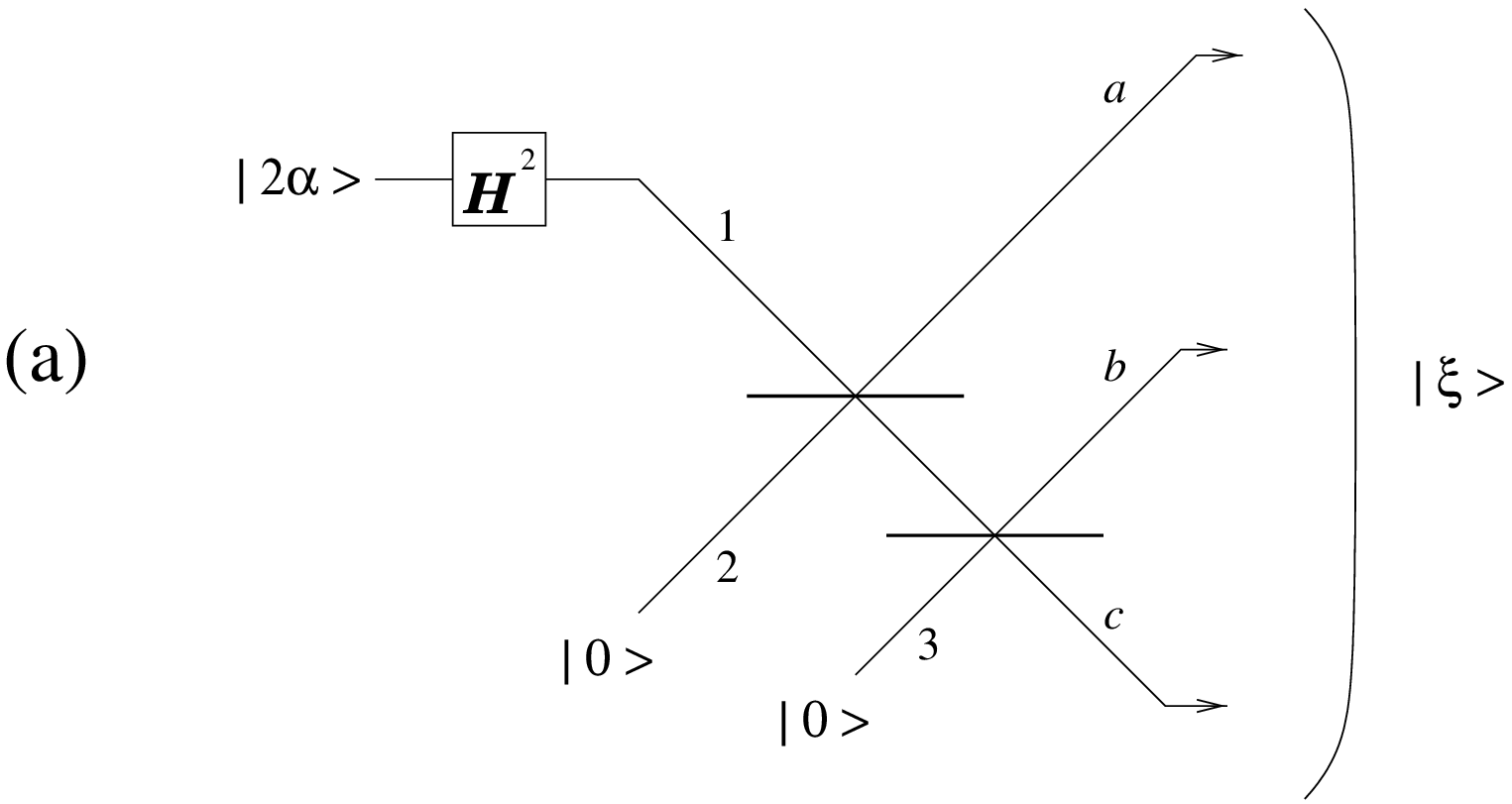}}}
\centerline{\scalebox{0.5}{\includegraphics{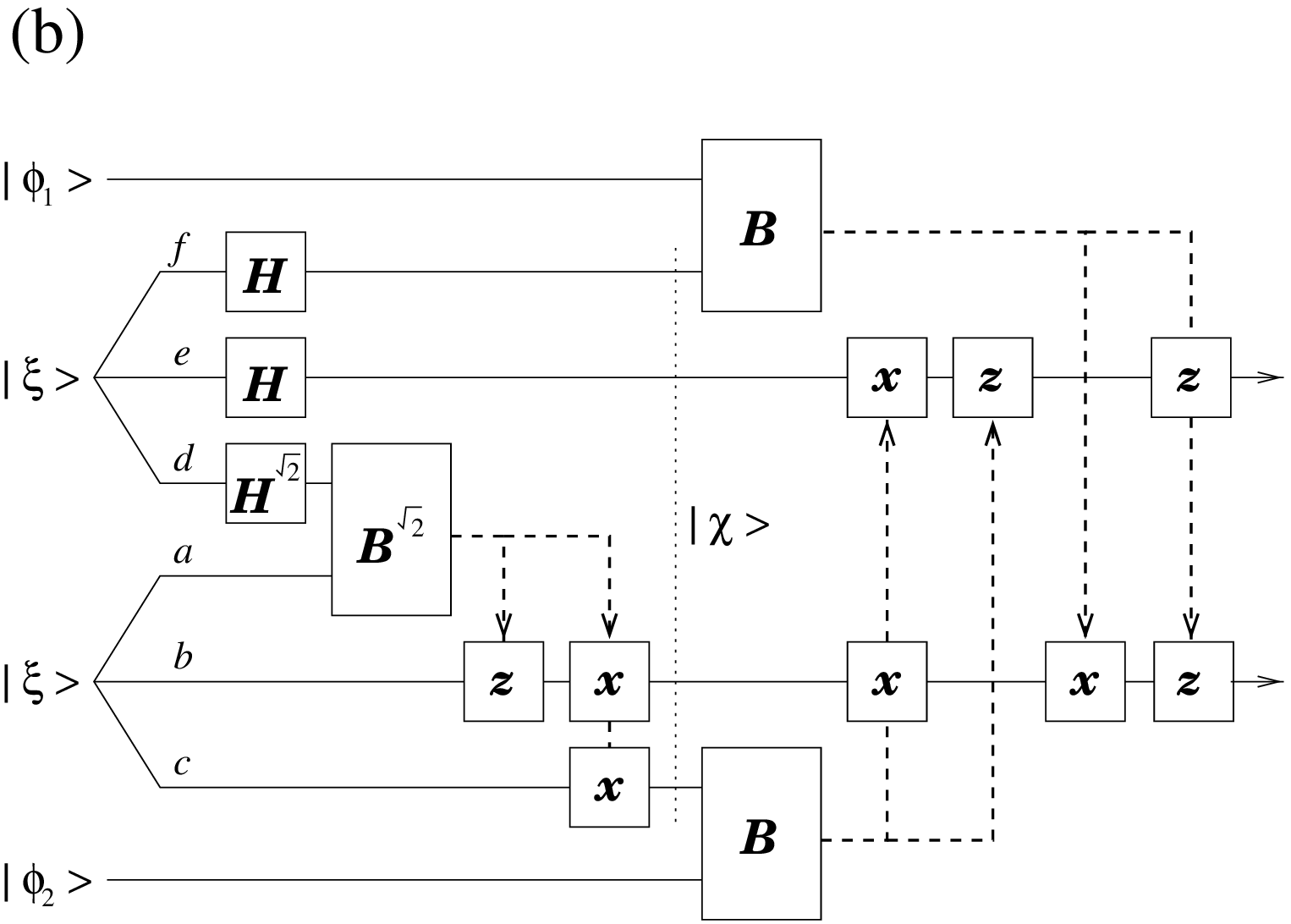}}}
\vspace{0.5cm}
\caption{CNOT operation using teleportation protocol
  and three-mode entanglement. (a) Generation of a three-mode
  entangled state $|\xi\rangle$ = ${\cal N}(|\sqrt{2}\alpha,\alpha,\alpha\rangle$ +
  $|-\sqrt{2}\alpha,-\alpha,-\alpha\rangle$ with beam splitters.
  $H^2$-gate is the Hadamard gate with an incident qubit state of
  amplitudes $\pm2\alpha$.  (b) CNOT operation with the use of the
  coherent field $|\xi\rangle$ and the teleportation protocol. A
  four-mode entangled state $|\chi\rangle$ is generated for the
  operation at the left-hand side of the circuit.  $|\phi_1\rangle$ is
  the target bit and $|\phi_2\rangle$ is the control bit here.}
\label{fig:cnot}
\end{figure}

For any ideal teleportation scheme, a maximally entangled pair, Bell
measurement and unitary operations are required \cite{Bennett93}.  In
our case, necessary unitary operations $\sigma_x$ and $\sigma_z$
correspond to a phase shift $P(\pi)$ and displacement operation
$D(\frac{i\pi}{4\alpha\sqrt{1-T}})$ respectively.  An entangled
coherent channel $|\Phi_+\rangle={\cal
  N}_+(|\alpha\rangle|\alpha\rangle+|-\alpha\rangle|-\alpha\rangle)$,
where ${\cal N}_+$ is a normalization factor, can be generated from a
coherent state passing through a $H^{\sqrt{2}}$ gate and a 50-50 beam
splitter as shown in Fig.~\ref{fig:tc}(a).
The superscript $\sqrt{2}$ in $H^{\sqrt 2}$ stands for the amplitude of
the incident field being $\sqrt{2}\alpha$.  Note that the coherent
amplitude $i\Delta$ for a unitary operation shown in Fig.~\ref{fig:h}
should be $i\pi/[8\alpha\sqrt{2(1-T)}]$ for the $H^{\sqrt 2}$-gate
operation.  The Bell measurement shown in Fig.~\ref{fig:tc}(b) is to
distinguish four quasi-Bell states \cite{Hirota01},
\begin{eqnarray}
\label{eq:qbs1}
|\Phi_\pm\rangle&=&{\cal N}_\pm(|\alpha,\alpha\rangle\pm|-\alpha,-\alpha\rangle),\\
\label{eq:qbs2}
|\Psi_\pm\rangle&=&{\cal N}_\pm(|\alpha,-\alpha\rangle\pm|-\alpha,\alpha\rangle),
\end{eqnarray}
where
$|\pm\alpha,\pm\alpha\rangle=|\pm\alpha\rangle\otimes|\pm\alpha\rangle$.
  Note that the quasi-Bell states become
maximally entangled Bell states when $\alpha$ is large. 
If the incident field to the first beam splitter in Fig.~\ref{fig:tc}
(b) is $|\Phi_+\rangle_{12}$, it becomes
$|0,2\alpha,-\sqrt{2}\alpha,\sqrt{2}\alpha\rangle_{abcd}$ at detectors
$A$, $B$, $C$, and $D$. If detector $A$ does not click while the others
do, the measurement outcome is $|\Phi_+\rangle_{12}$.  Likewise, only
$B$ does not click for the measurement outcome $|\Phi_-\rangle_{12}$,
$C$ for $|\Psi_+\rangle_{12}$, and $D$ for $|\Psi_-\rangle_{12}$.  The
failure probability for which no photon is detected at more than one
detector, which is due to the non-zero probability of $\langle 0|\pm
2\alpha\rangle$ and $\langle 0|\pm\sqrt{2}\alpha\rangle$, approaches
to zero rapidly as $\alpha$ increases, and, moreover, the failure is
always known when it occurs.  The scheme to teleport $|\phi\rangle$
via the entangled channel $|\Phi_+\rangle$ is summarized in
Fig.~\ref{fig:tc}(c).  When the Bell measurement outcome is
$|\Phi_+\rangle$, the output state does not need any operation. 
When the Bell measurement outcome is
$|\Phi_-\rangle$ or $|\Psi_+\rangle$,
$\sigma_z$ or $\sigma_x$
is required respectively.
The unitary operations $\sigma_z$ and $\sigma_x$ should be successively
applied for the outcome $|\Psi_-\rangle$.

Gottesman and Chuang showed that the teleportation protocol can be
used to construct a CNOT gate 
\cite{GC}.  To apply their suggestion in our scheme, we
need to use two three-mode entangled states represented by
\begin{equation}
\label{eq:three}
|\xi\rangle ={\cal N} \Big(|\sqrt{2}\alpha,\alpha,\alpha\rangle +
|-\sqrt{2}\alpha,-\alpha,-\alpha\rangle\Big),
\end{equation}
where ${\cal N}$ is a normalization factor, and the quantum
teleportation protocol we just developed. The entangled state
$|\xi\rangle$ can be generated by passing a coherent field
$|2\alpha\rangle$ through a $H^2$-gate, which is a Hadamard gate for a
qubit with logical bases $|\pm2\alpha\rangle$, and two 50-50 beam
splitters as shown in Fig.~\ref{fig:cnot}(a).  After generating
$|\xi\rangle_{abc}$ and $|\xi\rangle_{def}$, Hadamard operations are
applied to $|\xi\rangle_{def}$ as shown in Fig.~\ref{fig:cnot}(a).
This makes the given state $|\xi\rangle_{abc}\otimes|\xi\rangle_{def}$
to be
\begin{eqnarray}
&&~~\longrightarrow
|\Phi_+^\prime\rangle_{ad}\Big\{|\alpha,\alpha\rangle\big(|\alpha,\alpha\rangle +
|-\alpha,-\alpha\rangle\big)\nonumber\\
&&~~~~~~~~~~~~~~~~~+|-\alpha,-\alpha\rangle\big(|\alpha,-\alpha\rangle +
|-\alpha,\alpha\rangle\big)\Big\}_{bcef}\nonumber\\
&&~~~~~~~~|\Phi^\prime_-\rangle_{ad}\Big\{|\alpha,\alpha\rangle\big(|\alpha,\alpha\rangle +
|-\alpha,-\alpha\rangle\big)\nonumber\\
&&~~~~~~~~~~~~~~~~~-|-\alpha,-\alpha\rangle\big(|\alpha,-\alpha\rangle +
|-\alpha,\alpha\rangle\big)\Big\}_{bcef}\nonumber\\
&&~~~~~~~~|\Psi^\prime_+\rangle_{ad}\Big\{|-\alpha,-\alpha\rangle\big(|\alpha,\alpha\rangle +
|-\alpha,-\alpha\rangle\big)\nonumber\\
&&~~~~~~~~~~~~~~~~~+|\alpha,\alpha\rangle\big(|\alpha,-\alpha\rangle +
|-\alpha,\alpha\rangle\big)\Big\}_{bcef}\nonumber\\
&&~~~~~~~~|\Psi^\prime_-\rangle_{ad}\Big\{|-\alpha,-\alpha\rangle\big(|\alpha,\alpha\rangle +
|-\alpha,-\alpha\rangle\big)\nonumber\\
&&~~~~~~~~~~~~~~~~~-|\alpha,\alpha\rangle\big(|\alpha,-\alpha\rangle +
|-\alpha,\alpha\rangle\big)\Big\}_{bcef},
\label{long}
\end{eqnarray}
where $|\Phi^\prime_\pm\rangle$ and $|\Psi^\prime_\pm\rangle$ are quasi-Bell states
with the coherent amplitude $\pm\sqrt{2}\alpha$ and the  normalization
factor is omitted.  The Bell measurement
$B^{\sqrt 2}$ in the figure,
must be performed on modes $a$ and
$d$. 
It can be easily shown from Eq.~(\ref{long}) that 
 a four-mode entangled state 
\begin{eqnarray}
|\chi\rangle_{bcef}
 =&& {\cal N}^\prime\Big[|\alpha,\alpha\rangle\big(|\alpha,\alpha\rangle +
|-\alpha,-\alpha\rangle\big)\nonumber\\ &&~~~
+|-\alpha,-\alpha\rangle\big(|\alpha,-\alpha\rangle +
|-\alpha,\alpha\rangle\big)\Big],
\end{eqnarray}
where ${\cal N}^\prime$ is a normalization factor, is generated after
the appropriate unitary operation according to the Bell measurement
result as shown in Fig.~\ref{fig:cnot}(b).
The entangled state $|\chi\rangle_{bcef}$ is used to complete the CNOT
gate on the right-hand side of the circuit in Fig.~\ref{fig:cnot}(b),
which can be verified by a little algebra \cite{GC}.

\section{Estimation of possible errors}

We have shown that 
universal quantum computation using coherent states can be realized
using coherent states.
We already pointed out that the
failure probability for the measurement which is of the order of
$|\langle\sqrt{2}\alpha|0\rangle|^2$ is not only very small for
a reasonably large $\alpha$ but also the failure is known whenever it occurs.
If the detection efficiency of a photodetector is $d$, the failure
probability $P^d_f$ of the detector not to register any photon, while
the incident field is 
$|\phi_R\rangle_{ab}$ in Eq.~(\ref{eq:ax}), is
\begin{eqnarray}
P^d_f&=&\sum_{n,m=0}^{\infty}|_a\langle n |_b\langle m|\phi_R\rangle_{ab}|^2
(1-d)^n  (1-d)^m. \nonumber\\
&\approx&\sum_{n=0}^{\infty}|\langle n|\sqrt{2}\alpha\rangle|^2 (1-d)^n,
\end{eqnarray}
where approximation (\ref{nap}) is used.
For example, suppose that
 $\alpha=3$ and the detection
efficiency of the detectors is 90\% which is a reasonable value for an
avalanched photodetector \cite{Takeuchi}, 
the failure
probability $P^d_f$ that the detector misses all the photons is
$P_f^d\approx9\times10^{-8}$.

If the effect of $\ep$ for the displacement operator
is not negligible, a qubit state $|\phi^{\prime}\rangle_1=D(i\ep_1)\cdots
D(i\ep_N)|\phi\rangle_1$ after $N$ displacement operations
may be
\begin{equation}
\label{out}
|\phi^{\prime}\rangle_1={\cal A}^\prime\Big|\alpha+i\sum_{n=1}^N\ep_n\Big\rangle_1+{\cal B}^\prime\Big|-\alpha+i\sum_{n=1}^N\ep_n\Big\rangle_b.
\end{equation}
After passing a 50-50 beam splitter with an auxiliary state
$|\alpha\rangle_2$ as shown in Fig.~\ref{fig:readout}, the state $|\phi^\prime\rangle_1$ becomes
\begin{eqnarray}
\label{eq:ax2}
&&|\phi_R^\prime\rangle_{ab}={\cal A}^\prime\Big|\sqrt{2}\alpha+\frac{i}{\sqrt{2}}\sum_{n=1}^N\ep_n\Big\rangle_a\Big|\frac{i}{\sqrt{2}}\sum_{n=1}^N\ep_n\Big\rangle_b \nonumber\\
&&~~~~~~~~~~~~~+{\cal B}^\prime\Big|\frac{i}{\sqrt{2}}\sum_{n=1}^N\ep_n\Big\rangle_a\Big|-\sqrt{2}\alpha+\frac{i}{\sqrt{2}}\sum_{n=1}^N\ep_n\Big\rangle_b.
\end{eqnarray}
  In this condition,
there is non-zero probability $\widetilde P_f^d$ for
undetected errors in which detector $A(B)$ detects any photon and
$B(A)$ does not while the incident state $|\phi^\prime\rangle_1$ was
$|1_L\rangle$ ($|0_L\rangle$) (see Fig.~\ref{fig:readout}). For the
worst case, all $\ep_n$'s may have the same sign with a large $N$.
One useful trick to overcome this problem is to flip the sign of
$\ep_n$ appropriately for each operation, noting that the rotation
$R_z(\theta)$ can be performed both by positive and negative
$\theta$. By this way, we can keep
$\sum_{n=1}^N\ep_n\sim\bar\ep=\pi/4\alpha$, regardless
of $N$, then Eq.~(\ref{eq:ax2}) can be represented as
\begin{eqnarray}
\label{eq:ax3}
&&|\phi_R^\prime\rangle_{ab}={\cal A}^\prime\Big|\sqrt{2}\alpha+\frac{i{\bar \ep}}{\sqrt{2}}\Big\rangle_a\Big|\frac{i{\bar \ep}}{\sqrt{2}}\Big\rangle_b \nonumber\\
&&~~~~~~~~~~~~~+{\cal B}^\prime\Big|\frac{i{\bar \ep}}{\sqrt{2}}\Big\rangle_a\Big|-\sqrt{2}\alpha+\frac{i{\bar \ep}}{\sqrt{2}}\Big\rangle_b.
\end{eqnarray}
In this condition, the fidelity between the final state (\ref{out})
and the ideal output is proportional to $e^{-\ep^2}$ from
Eq.~(\ref{fidelity-rotation}). Fidelity of $\approx0.93$ is then obtained
for $\alpha=3$.

Differently from $P^d_f$, the undetected error probability $\widetilde
P_f^d$ is a probability of making an error without being recognized.
Considering the accumulated error as in Eq.~(\ref{eq:ax3}), in order
to minimize the undetected
error $\widetilde P_f^d$ while keeping $P_f^d$ low, we need to modify the
criterion to discriminate $|\pm\sqrt{2}\alpha+i\bar\ep/\sqrt{2}\rangle$ and
$|i\bar\ep/\sqrt{2}\rangle$.  Ideally we took $\bar \ep=0$ and discriminated
the two states by detection of any photons and no photon.
 In this case, the
probability of $|\pm\sqrt{2}\alpha+i\bar \ep/\sqrt{2}\rangle$ registering no
photon is
\begin{eqnarray}
p_A&=&\sum_{n=0}^{\infty}|\langle
n|\pm\sqrt{2}\alpha+
i{\bar\ep}/\sqrt{2}\rangle|^2(1-d)^n
\end{eqnarray}
and the probability of the state $|i\bar \ep/\sqrt{2}\rangle$
registering one or more photons is
\begin{eqnarray}
p_B&=&\sum_{m=1}^\infty\sum_{n=m}^{\infty}|\langle
n|i{\bar\ep}/\sqrt{2}\rangle|^2 {}_n
C_md^m(1-d)^{n-m} 
\end{eqnarray}
where $_nC_m=n!/m!(n-m)!$. Both $p_A$ and $p_B$ approach to zero as $\alpha$ increases.
We then obtain undetected error probability 
${\widetilde P_f^d}=p_A\times p_B$. 
On the other hand, the success probability $P_s$ is the  probability in that
$|i{\bar\ep}/\sqrt{2}\rangle$ yields no photon and $|\pm\sqrt{2}\alpha+i{\bar\ep}/\sqrt{2}\rangle$
yields any photon(s):
\begin{eqnarray}
&&P_s=\sum_{n=0}^\infty|\langle n|i\ep/\sqrt{2}\rangle|^2
(1-d)^n
\times\nonumber \\
&&\sum_{m=1}^{\infty}\sum_{n=m}^{\infty} |\langle n|\sqrt{2}\alpha+i{\bar\ep}/\sqrt{2}\rangle|^2 {}_n
C_md^m(1-d)^{n-m}.
\end{eqnarray}
The detected
 error probability
 is $P_f^d=1-P_s-\widetilde P_f^d$.
Suppose that $\alpha=3$ ($\bar \ep$ is then $\approx 0.26$), and the detection
efficiency is again 90\%,
$p_A\approx9\times10^{-8}$ and $p_B\approx0.030$ are obtained.  If we
keep the criterion for the ideal case, we find $\widetilde
P_f^d\approx3\times10^{-9}$ and $P_f^d\approx0.030$.
However, if we take the registration of 0,1 and 2 photons as the measurement of
$|i{\bar\ep}/\sqrt{2}\rangle$ then $p_A$, $p_B$ and $P_s$ should be
re-defined 
 as follows:
\begin{eqnarray}
p_A&=&\sum_{n=0}^{\infty}|\langle
n|\sqrt{2}\alpha+
i{\bar\ep}/\sqrt{2}\rangle|^2(1-d)^n\nonumber
\\&&+\sum_{n=1}^{\infty}|\langle
n|\sqrt{2}\alpha+
i{\bar\ep}/\sqrt{2}\rangle|^2d(1-d)^{n-1}\nonumber
\\&&+\sum_{n=2}^{\infty}|\langle
n|\sqrt{2}\alpha+
i{\bar\ep}/\sqrt{2}\rangle|^2d^2(1-d)^{n-2}\label{AA}
\\
p_B&=&\sum_{m=3}^\infty\sum_{n=m}^{\infty}|\langle
n|i{\bar\ep}/\sqrt{2}\rangle|^2 {}_n
C_md^m(1-d)^{n-m}\label{BB}\\
P_s&=&\bigg\{
\sum_{n=0}^{\infty}|\langle
n|i{\bar\ep}/\sqrt{2}\rangle|^2(1-d)^n\nonumber
\\&&+\sum_{n=1}^{\infty}|\langle
n|i{\bar\ep}/\sqrt{2}\rangle|^2d(1-d)^{n-1}\nonumber
\\&&+\sum_{n=2}^{\infty}|\langle
n|i{\bar\ep}/\sqrt{2}\rangle|^2d^2(1-d)^{n-2}\bigg\}\times
\sum_{m=3}^\infty\sum_{n=m}^{\infty}\nonumber
\\
&&|\langle
n|\sqrt{2}\alpha+i{\bar\ep}/\sqrt{2}\rangle|^2 {}_n
C_md^m(1-d)^{n-m}.\label{CC}
\end{eqnarray}
We then find $\widetilde P_f^d\approx6\times10^{-11}$ and
$P_f^d\approx2\times10^{-5}$
for $\alpha=3$ and $d=0.9$.
Recently, Takeuchi {\em et al.} \cite{Takeuchi} developed an
avalanched photodetector which can discern 0,1, and 2 photons with
high efficiency.

Decoherence is considered one of the main obstacles in quantum computation. 
When a qubit state $|\phi\rangle$ is
subject to a vacuum environment it evolves to \cite{Phoenix}
\begin{eqnarray} 
\nonumber
&&\rho_M(\tau)={{\cal N}_\tau}\bigg\{
|{\cal A}|^2|t\alpha\rangle\langle t\alpha|+
|{\cal B}|^2|-t\alpha\rangle\langle -t\alpha|\\
&&~~~~~~~~~~~~~~+\Gamma\Big({\cal AB^*}|t\alpha\rangle\langle -t\alpha|+
{\cal A^*B}|-t\alpha\rangle\langle t\alpha|\Big)\bigg\}
\end{eqnarray}
where $t=e^{-\gamma \tau/2}$, $\Gamma=e^{-2(1-t^2)\alpha^2}$, $\gamma$
is the 
energy decay rate, $\tau$ is the interaction time, and ${\cal
  N}_\tau$ is the normalization factor.  Considering decoherence, we need
to change $|0_L\rangle$ and $|1_L\rangle$ to $|t\alpha\rangle$ and
$|-t\alpha\rangle$.  The auxiliary coherent fields for computation
have to be changed likewise.  The larger the initial coherent
amplitude $\alpha$ is, the longer the condition that $\langle
t\alpha|-t\alpha\rangle\approx0$ is preserved, but the decoherence
becomes more rapid as $\alpha$ increases because $\Gamma$ decreases
more rapidly for a larger $\alpha$.
  The energy decay rate $\gamma$ of the relevant
system and number of required operations for computation may be the
crucial factors to decide the value of $\alpha$.
 However, decohered states can
still be represented by combinations of 1-bit errors for
time-dependent logical qubits $|t\alpha\rangle$ and
$|-t\alpha\rangle$.  It is known that an error correction circuit for
an arbitrary 1-qubit error can be built using CNOT and 1-bit unitary
operations \cite{L96}. 

\section{remarks}
In conclusion, we have found that near-deterministic universal 
quantum computation can be realized using coherent states.  Efficient
readout is possible using beam splitters and coherent light sources.
Single-bit unitary transformation can be performed using beam
splitters and nonlinear media, and CNOT gate can be constructed based
on teleportation protocol.  Teleportation of a coherent state qubit
can be accomplished with a complete Bell measurement for a large
coherent amplitude using nonlinear media, photodetectors, coherent
light sources, and beam splitters. 
Decohered states can be represented by combinations of 1-bit errors
for time-dependent coherent state qubits of reduced amplitude. 
A purification scheme for decohered entangled channels 
 has been studied \cite{JKpurify}.
Detailed error correction methods for our scheme deserves further
investigation.  The nonlinear effect \cite{Yurke} used in this paper
is typically too weak to generate required superposition states in
current technology.  The study of generating coherent superposition of
optical states requires further study.

\acknowledgements

We thank the UK Engineering and Physical Sciences Research Council for
financial support through GR/R33304. HJ acknowledges the Overseas Research 
Student award.

\end{document}